# A tentative derivation of the main cosmological parameters

**A. Dinculescu** [1]

**Abstract**     Based on the assumption that some apparent properties of the observable universe are accurate at a reasonable level of approximation, a tentative is made to independently derive the values of the baryon density parameter, the Hubble constant, the cosmic microwave background temperature and the helium mass fraction. The obtained values are in excellent agreement with those given by the most recent observational data.

**Keywords**     Cosmology – Observable universe – Cosmological parameters

―――――

e-mail: *adinculescu@cox.net*

# 1. Introduction

Cosmological parameters lay at the very foundation of the standard cosmological model. The finding of their approximate values took years of hard work and painstaking measurements that are still continuing in order to reduce the existing margin of error. It would be therefore useful if one could find a way to independently derive or at least approximate these values. Based on previous works of Weyl (1919), Eddington (1923), Dirac (1938), Weinberg (1972) and others, which have shown that some characteristics of the observable universe can be expressed as functions of fundamental constants of nature, it was suggested (Dinculescu 2005) that a possible explanation for the existence of these characteristics is the close similarity of the observable universe to a precisely defined and apparently immovable entity called "the photon mean free path sphere". Here we use this similarity and recent observations to derive the values of four of the most important cosmological parameters: the baryon density parameter $\Omega_b$, the Hubble constant $H$, the temperature $T_{CMB}$ of the cosmic microwave background ($CMB$), and the helium mass fraction $Y_P$. At the base of our derivation there is the assumption that some apparent properties of the observable universe related to its baryonic mass $M_b$, its size $R_H$, or its radiation energy $E_{CMB}$ are accurate at a reasonable level of approximation, and can be used as premises for the calculation of its parameters. Because the observable universe is sometimes called the speed of light sphere (Ellis and Rothman 1993) or the Hubble sphere (Harrison 1991) here we adopt the term *"speed of light sphere"* (*SLS*) to highlight a possible connection with the photon mean free path sphere defined below.

## 2.    The premises

Let us consider a large enough volume of space containing a fully ionized, homogenous mixture of matter and radiation. The matter is characterized by the baryon density $\rho_b = 3\Omega_b H^2 / 8\pi G$ (where $G$ is the gravitational constant), the baryon number density $n_b = \rho_b / m_p$ (where $m_p$ is the mass of the proton), the helium mass fraction $Y_p = M_{He} / M_b$ (where $M_{He}$ is the total mass of helium) and the electron number density $n_e = \left(1 - Y_p / 2\right)n_b$. The radiation is characterized by the temperature $T_\gamma$ and the energy density $u_\gamma = a T_\gamma^4$, where $a$ is the radiation density constant. We define the photon mean free path sphere (*ph.m.f.p.s.*) in the above space as the sphere of radius $R_{ph}$ that has an optical depth $\tau_{ph} = R_{ph}\,\sigma_T\,n_e$ equal to unity. Here $\sigma_T = \left(8\pi / 3\right)r_e^2$ is the Thomson cross section, $r_e = e^2 / m_e c^2$ is the classical radius of the electron, $e$ is the elementary electric charge in Gaussian units, $m_e$ is the mass of the electron, and $c$ is the speed of light.  It is easy to see that the radius of a *ph.m.f.p.s.* depends only on the number density of the electrons inside and practically can take any value. However, if the *ph.m.f.p.s.* is self-gravitating its radius cannot be larger than a certain value $R_{Ph}$ above which its mass $M_{Ph}$ collapses into a black hole. (Note the capital "$P$" in the subscript to differentiate it from the $ph.m.f.p.s.$). This maximal photon mean free path sphere ($MxPhMFPS$) has several characteristics of interest. Thus, from

$$R_{Ph}\,\rho_{Ph} = \frac{m_p}{\left(1 - Y_p / 2\right)\sigma_T} \tag{1}$$

one has

$$M_{Ph} = \frac{4\pi}{3}\,\rho_{Ph}\,R_{Ph}^3 = \frac{1}{2}\,\frac{m_p}{1 - Y_p / 2}\left(\frac{R_{Ph}}{r_e}\right)^2 \tag{2}$$

Since $n_e = \left(1 - Y_p / 2\right)n_b$, the number of protons and electrons in any $ph.m.f.p.s.$ is

$$N_{(p+e)} = \left(\frac{R_{ph}}{r_e}\right)^2 \tag{3}$$

(Note the similarity with Eddington's "number of particles in the universe").

With $2\,G\,M_{Ph} / R_{ph}\,c^2 = 1$ by definition, (2) gives

$$R_{Ph} = \left(1 - Y_p / 2\right)N_{p,e}\,r_e \tag{4}$$

where

$$N_{p,e} = \frac{e^2}{G\,m_p\,m_e} \tag{5}$$

is the ratio of electric to gravitational forces between the proton and the electron. (Note the similarity with Dirac's "large number"). It follows that

$$M_{Ph} = \frac{1 - Y_p / 2}{2}\,N_{p,e}^2\,m_p \tag{6}$$

Since the number density of protons or electrons is smaller than the baryon number density by a factor of $1 - Y_p / 2$, all *ph.m.f.p.s.* have the same surface density of similar particles $i$ ( protons or electrons):

$$\frac{M_{ph(i)}}{R_{ph}^2} = \frac{1}{2} \frac{m_i}{r_e^2} \qquad (7)$$

Consequently, the energy density corresponding to gravitational interactions between similar particles is the same in all *ph.m.f.p.s.*

$$u_{ph(i)} = \frac{3}{5} \frac{G M_{ph(i)}^2}{\frac{4\pi}{3} R_{ph(i)}^4} = \frac{4\pi}{5} \frac{G m_i^2}{\sigma_T^2} \qquad (8)$$

If all neutrons in the *MxPhMFPS* undergo β-decay, what remains are only protons and electrons. Their total number is

$$N_{Ph(tot)} = 2 \frac{M_{Ph}}{m_p} = \left(1 - Y_p / 2\right) N_{p,e}^2 \qquad (9)$$

*Observation no. 1:* Some theoretical considerations (Dinculescu 2009) point to an apparent equality between the baryonic mass of the *SLS* and the mass of a *MxPhMFPS* with the same composition.

**Premise no. 1**: *The baryonic mass of the SLS equals the mass of a MxPhMFPS that has the same composition as the SLS.*

One has $M_b = M_{Ph}$. With $2 G M_b / R_H c^2 = \Omega_b$ and $2 G M_{Ph} / R_{ph} c^2 = 1$ one obtains

$$R_H = \frac{R_{Ph}}{\Omega_b} \qquad (10)$$

*Observation no. 2:* From all possible pairs of known elementary particles, the only particles that fit the Dirac type equation

$$\frac{R_H}{\left(\lambda_i \, \lambda_j\right)^{1/2}} = \frac{e^2}{G \, m_i \, m_j} \qquad (11)$$

are the proton and the electron (Dinculescu 2006). Here $\lambda_{i(j)}$ is the Compton wavelength corresponding to the particle $i(j)$.

**Premise no. 2:** *The size of the SLS is related to the Compton wavelengths of its constituent particles by*

$$R_H = N_{p,e} \, \lambda_{p,e} \qquad (12)$$

where $\lambda_{p,e} = \left(\lambda_p \, \lambda_e\right)^{1/2}$ is the geometric mean between the Compton wavelength of the proton $\lambda_p$ and the Compton wavelength of the electron $\lambda_e$.

*Observation no. 3:* It appears that the radiation energy per particle in a *SLS* containing only protons and electrons is on the order of the energy per particle released in the process of electron – positron annihilation (Dinculescu 2009).

**Premise no. 3:** *The radiation energy per particle in a SLS containing only protons and electrons equals the rest energy of the electron.* One has

$$E_{CMB} = N_{tot} \, m_e \, c^2 \qquad (13)$$

With the total number of particles given in this case by $N_{tot} = N_{Ph(tot)} = 2 \, N_b$, where $N_b$ is the total number of baryons, the radiation energy density in the *SLS* is

$$a \, T_{CMB}^4 = n_{tot} \, m_e \, c^2 \qquad (14)$$

*Observation no. 4:* Since the time derivative $dE/dt$ of any amount of energy cannot exceed $c^5/G$ and the *CMB* was once by far the most dominating form of energy in the universe, the luminosity of the source at its origin was probably close to this value (Dinculescu 2007a).

**Premise no. 4**: *The radiation energy in the SLS corresponds to the luminosity of a system that released a quantity of energy equivalent to the maximum energy in baryonic matter in the minimum time allowed by causality.*

One has

$$L_{CMB} = \frac{1}{2} \frac{M_b\, c^2}{R_H / c} \tag{15}$$

With $2\, G\, M_b / R_H\, c^2 = \Omega_b$ one obtains

$$L_{CMB} = \frac{\Omega_b}{4} \frac{c^5}{G} \tag{16}$$

As known, the radiation energy inside a spherical volume of space of radius $R$ due to a source of luminosity $L$ at its center is $E = L\, t_{esc}$ , where $t_{esc} = \sigma_T\, n_e\, R^2 / c$ is the escape time of a photon from the sphere. If the system contains only protons and electrons, then $n_e = n_p = n_b = n_{tot} / 2$ and

$$E_{CMB} = \frac{4\,\pi}{3} \frac{L_{CMB}\, r_e^2}{c}\, n_{tot}\, R_H^2 \tag{17}$$

It follows that

$$a\, T_{CMB}^4 = \frac{L_{CMB}\, r_e^2}{c} \frac{n_{tot}}{R_H} \tag{18}$$

Since a source of radiation exerts pressure on the surrounding matter, one can think of it as *a "radiation charge"*. As in the case of assembling a sphere of uniformly distributed electric charges, one has to spend energy in order to assemble a spherical cloud of radius $r$ around a radiation source of luminosity $L$. The radiation force on a spherical shell of radius $r$ and thickness $dr$ is

$$dF_\gamma = \frac{L}{c}\, \sigma_T\, n_e\, dr \qquad (19)$$

The energy required to assemble the entire sphere is

$$E_\gamma = \int_0^r r\, dF_\gamma = \frac{1}{2}\frac{L}{c}\, \sigma_T\, n_e\, r^2 \qquad (20)$$

Therefore, one can define the radiation charge as $Q_\gamma \equiv \left(2\, E_\gamma\, r\right)^{1/2}$ (where the factor of two takes account of the fact that unlike the interaction between electric charges which is reciprocal, the force exerted by the radiation charge does not have an equal counterpart from the irradiated particles) and write

$$a\, T_{CMB}^4 = \frac{3}{4\pi}\frac{Q_{CMB}^2}{R_H^4} \qquad (21)$$

With $N_{tot} = 2\, N_e$ , the radiation charge per particle $q_\gamma = Q_\gamma / N_{tot}$ is given by

$$q_\gamma^2 = \frac{L}{c\, N_{tot}}\, r_e^2 \qquad (22)$$

## 3.    The derivation

When one equates (14) with (18) using (16), one obtains

$$\Omega_b = 2\left[\left(1 - Y_P / 2\right)\frac{m_e}{m_p}\right]^{1/2} \qquad (23)$$

On equating this with

$$\Omega_b = \left(1 - Y_P / 2\right)\frac{r_e}{\lambda_{p,e}} \qquad (24)$$

obtained from (4), (10) and (12), one arrives at

$$1 - Y_P / 2 = 4 \left( \frac{\lambda_{p,e}}{r_e} \right)^2 \frac{m_e}{m_p} \qquad (25)$$

and

$$\Omega_b = 4 \frac{\lambda_{p,e}}{r_e} \frac{m_e}{m_p} \qquad (26)$$

These two equations give $Y_P = 0.2409$, vs. $Y_P = 0.2477 \pm 0.0029$ (Peimbert et al. 2007), and $\Omega_b = 0.0438$ vs. $\Omega_b = 0.0456 \pm 0.0015$ (Komatsu, et al. 2009).

With $R_H = c / H$, equation (12) reads

$$H = \frac{c}{N_{p,e} \lambda_{p,e}} \qquad (27)$$

This gives for the Hubble constant a calculated value $H = 71.98 \, km \, s^{-1} Mpc^{-1}$ vs. $H = 70.5 \pm 1.3 \, km \, s^{-1} Mpc^{-1}$ (Komatsu, et al. 2009). It follows that $\Omega_b \, h^2 = 0.02268$ (where $h = H / 100 \, km \, s^{-1} Mpc^{-1}$) vs. $\Omega_b \, h^2 = 0.02267 \pm 0.00059$ (Komatsu, et al. 2009). With

$$n_b = \frac{3}{8 \pi} \frac{\Omega_b \, c^2}{G \, m_p \, R_H^2} \qquad (28)$$

using (12) and (26) in (14), (18), or (21), one can write the *CMB* energy density as a function of universal constants only. One has

$$a \, T_{CMB}^4 = \frac{3}{\pi} \frac{G \, m_e^2}{r_e^3 \, \lambda_{p,e}} \qquad (29)$$

This gives for the *CMB* temperature a calculated value $T_{CMB} = 2.7254 \, K$ vs. $T_{CMB} = 2.725 \pm 0.001 \, K$ (Fixen and Mather 2002).

From the above equations one can derive other relationships between atomic and cosmological quantities. Thus, when one equates (14) with (18) using (15), one obtains

$$\frac{M_b}{R_H^2} = 2\,\frac{m_e}{r_e^2} \tag{30}$$

This says the surface density of the *SLS* is on the order of the electron surface density of all *ph.m.f.p.s.* (see 7). We note in passing that this is also on the order of the surface density of the dark matter halos in galaxies (Martins 2009).

When one equates the radiation force $dF_\gamma$ (see 19) acting on a spherical shell of radius $R_H$ and thickness $dr$ with the corresponding gravitational force $dF_G = 4\pi\,G\,M_b\,m_p\,n_b\,dr$ and, in accordance with our premise no. 1, replaces $M_b$ by $M_{Ph}$ in the resulted Eddington luminosity

$$L_E = \frac{3}{2}\,\frac{G\,M_b\,m_p\,c}{r_e^2\left(1 - Y_p/2\right)} \tag{31}$$

one obtains

$$L_E = \frac{3}{4}\,\frac{c^5}{G} \tag{32}$$

This says the Eddington luminosity of the *SLS* is constant and equals the Eddington luminosities of all *MxPhMFPS* regardless of their composition. In connection with (16) this leads to

$$L_{CMB} = \frac{\Omega_b}{3}\,L_E \tag{33}$$

From (22), using (9), (16) and (24), one obtains

$$q_{CMB}^2 = \frac{G\,m_p\,m_e}{\Omega_b} \tag{34}$$

Hence, the $CMB$ radiation charge per particle is on the order of the gravitational charge of the proton and the electron.

## 5.    Concluding remarks

We wrote this paper being convinced that science can only benefit when one looks at the same problem from a different perspective, but the fact that we managed to derive the seemingly correct values of four of the most important cosmological parameters does not contradict in any way the current theory. Although our observations are probably correct, our premises can be wrong, at least in the present, rudimentary, form. However, due to the close match of our results with the observational data, they cannot be too wrong. Also, they contain some elements that in our opinion deserve to be more thoroughly analyzed. It is for example intriguing that, although we worked outside of the standard model, our equations (17), (18), (21), and (32), when applied to a commoving sphere of the size of the $SLS$ today, are independent of the redshift and hence valid at any epoch. This can be easily seen from the fact that the number of particles in a comoving sphere varies with the redshift $z$ as $(1+z)^3$, while its radius and the radiation temperature varies as $(1+z)^{-1}$ and $(1+z)$ respectively. Note that because in the $SLS$ the term $n_b R_H^2$ is constant, the radiation energy inside the $SLS$ according to (17) does not change with the redshift. It is this superposition of the $SLS$ and our commoving sphere at the present epoch that caused, in our opinion, some of the "numerical coincidences" reported in the literature, and allowed us to equate (14) with (18), although only (18) is redshift independent.

Another aspect that deserves to be more profoundly analyzed is the quasi equality between the radiation energy per particle in the *SLS* and the rest energy of the electron (a characteristic that certainly existed at the epoch of electron-positron annihilation). Is this a simple coincidence, or has a deeper meaning? At this moment we simply do not know. Of course, we can invoke the anthropic principle (Barrow and Tipler 1986), but this principle is too vague and hard if not impossible to verify. Or we can simply dismiss all these results, as precise as they are, as numerical coincidences, but they are too many, and too interconnected. Usually, when one encounters several numerical coincidences when trying to solve a problem, they are disparate and do not correlate to each other. In this case, the correlation is obvious. All our premises were independent, and still they complemented each other surprisingly well. As a proof of this interconnection are the relationships in (30), (32) and (34), which came out unexpectedly from our premises.

As a known cosmologist put it not long time ago, it is reasonable to expect that some of these curiosities are nothing more than accidents, but it is sensible to be aware of the possibility that some are clues to improvements in physics (Peebles 2003).